\documentclass[12pt,a4paper]{article}
\usepackage{amsmath}
\usepackage{latexsym}
\makeatletter
\def\rddots{\mathinner{\mkern1mu\raise\p@%
    \vbox{\kern7\p@\hbox{.}}\mkern2mu%
    \raise4\p@\hbox{.}\mkern2mu\raise7\p@\hbox{.}\mkern1mu}}
\makeatother
\newcommand{\fukuso}{{\mathbf C}}
\newcommand{\real}{{\mathbf R}}

\begin{document}

\title{\sl Canonical form of the Evolution Operator of 
a Time--Dependent Hamiltonian in the Three Level System}
\author{
  Kazuyuki FUJII
  \thanks{E-mail address : fujii@yokohama-cu.ac.jp }\\
  Department of Mathematical Sciences\\
  Yokohama City University\\
  Yokohama, 236--0027\\
  Japan
  }
\date{}
\maketitle
\begin{abstract}
  In this paper we study the evolution operator of 
a time--dependent Hamiltonian in the three level system. 
The evolution operator is based on $SU(3)$ and its dimension 
is $8$, so we obtain three complex Riccati differential equations 
interacting with one another (which have been obtained by Fujii 
and Oike) and two real phase equations. This is a canonical form 
of the evolution operator.
\end{abstract}

\section{Introduction}
In this paper we treat a finite dimensional quantum model and 
study the evolution operator of a time--dependent Hamiltonian 
from a geometrical point of view. 
Then we must solve the time--dependent Schr\"{o}dinger 
equation, which is a very hard task. Our method is based on 
one in \cite{RU-1}, \cite{RU-2} and \cite{FO-1}.

Let us start by setting the problem. By $H_{0}(n;\fukuso)$ we show 
the set of all $n\times n$--hermitian matrices with zero trace. 
The general form of $H$ is given by
\begin{eqnarray}
\label{eq:Hamiltonian}
&&H=H(t)
=
\left(
\begin{array}{cccccc}
h_{1}(t) & \bar{v}_{21}(t) & \bar{v}_{31}(t) & \cdots & \bar{v}_{n-1,1}(t) & \bar{v}_{n1}(t)  \\
v_{21}(t) & h_{2}(t) & \bar{v}_{32}(t) & \cdots & \bar{v}_{n-1,2}(t) & \bar{v}_{n2}(t)          \\
v_{31}(t) & v_{32}(t) & h_{3}(t) & \cdots & \bar{v}_{n-1,3}(t) & \bar{v}_{n3}(t)                  \\
\vdots & \vdots & \vdots & \ddots & \vdots  & \vdots                                          \\
v_{n-1,1}(t) & v_{n-1,2}(t) & v_{n-1,3}(t) & \cdots & h_{n-1}(t) & \bar{v}_{n,n-1}(t)           \\
v_{n1}(t) & v_{n2}(t) & v_{n3}(t) & \cdots & v_{n,n-1}(t) & h_{n}(t)
\end{array}
\right)\ \in\ H_{0}(n;\fukuso)  \nonumber \\
&&  \\
&&\qquad\qquad\qquad \mbox{with}\quad 
h_{1}(t)+h_{2}(t)+h_{3}(t)+\cdots +h_{n-1}(t)+h_{n}(t)=0. \nonumber
\end{eqnarray}
Under this Hamiltonian we want to find the evolution operator 
$U=U(t)$ satisfying the Schr\"{o}dinger equation
\begin{equation}
\label{eq:evolution-equation}
i\dot{U}\equiv i\frac{dU}{dt}=HU
\end{equation}
where we have set $\hbar=1$ for simplicity. The wave function 
$\Psi=\Psi(t)$ is of course related to the evolution operator like 
$\Psi(t)=U(t)\Psi(0),\ U(0)=E\ (: \mbox{identity})$. However, it is 
almost impossible to solve (\ref{eq:evolution-equation}) {\bf exactly} 
except for a few examples.

We decompose $U(t)$ into two unitary parts
\begin{equation}
\label{eq:decomposition}
U(t)=U_{1}(Z(t))U_{2}(t)
\end{equation}
where $Z$ is a local coordinate of some symmetric space like 
Grassmann manifolds or (generalized) flag manifolds, see for example 
\cite{Oi}, \cite{Fu-1}, \cite{Pi} and \cite{FO-2}. 
In the case of flag manifolds this decomposition is based on 
the approximation 
\begin{equation}
\label{eq:flag-manifold}
SU(n)\approx \left\{SU(n)/U(1)^{n-1}\right\} \times U(1)^{n-1}.
\end{equation}

In the following sections we treat 
the Bloch sphere $SU(2)/U(1)\cong {\fukuso}P^{1}\cong S^{2}$ and 
the flag manifold $SU(3)/U(1)^{2}$ as interesting examples. 

$U_{2}(t)$ is a small unitary part and in the paper it is set as 
the phase part
\begin{equation}
\label{eq:phase-part}
U_{2}(t)=
\left(
\begin{array}{cccc}
e^{i\phi_{1}(t)}       &            &                    \\
   & e^{i\phi_{2}(t)} &            &                    \\
   &                    & \ddots  &                    \\
   &                    &            & e^{i\phi_{n}(t)} 
\end{array}
\right);\quad \phi_{1}(t)+\phi_{2}(t)+\cdots +\phi_{n}(t)=0.
\end{equation}

Then the equation (\ref{eq:evolution-equation}) is reduced to 
a complicated combination of equations on $Z(t)=(z_{ij}(t))$ and 
$\{\phi_{k}(t)\}$. We want to call it a {\bf canonical form} in our sense. 
Unfortunately, it is very hard to write down the general case explicitly. 

In this paper we transform the equation (\ref{eq:evolution-equation}) 
into a canonical form in the case of $n=2$ and $n=3$. 
Namely, since $\dim_{\real}SU(2)$=3 for $n=2$ we obtain one complex 
Riccati differential equation and one real phase equation. See for example 
\cite{RU-1} and its references. 

On the other hand, since $\dim_{\real}SU(3)$=8 for $n=3$ we obtain three 
complex Riccati differential equations interacting with one another (which 
have been obtained by Fujii and Oike \cite{FO-2}) and two real phase equations.

\section{Two Level System}
In this section we treat the two level system ($n=2$) in detail. 
The Hamiltonian is
\begin{equation}
\label{eq:2-Hamiltonian}
H=
\left(
\begin{array}{cc}
h(t) & \bar{v}(t) \\
v(t) & -h(t)
\end{array}
\right)
\end{equation}
from (\ref{eq:Hamiltonian}). From the fact
\[
SU(2)\approx \frac{SU(2)}{U(1)}\times U(1)\cong S^{2}\times U(1)
\]
(see \cite{FO-2}), $U(t)$ can be parametrized as
\begin{equation}
\label{eq:2-decomposition}
U(t)=U_{1}(Z(t))U_{2}(t)
=
\frac{1}{\sqrt{1+|z(t)|^{2}}}
\left(
\begin{array}{cc}
1    & -\bar{z}(t) \\
z(t) & 1
\end{array}
\right)
\left(
\begin{array}{cc}
e^{i\phi(t)} &                  \\
               & e^{-i\phi(t)}
\end{array}
\right).
\end{equation}

From the equation (\ref{eq:evolution-equation}) we have only to 
calculate
\begin{equation}
\label{eq:2-evolution-equation}
H=i\dot{U}U^{\dagger}=i\dot{U}U^{-1}.
\end{equation}

Some calculation gives
\begin{eqnarray*}
\dot{U}U^{\dagger}
&=&
-\frac{\dot{z}\bar{z}+z\dot{\bar{z}}}{2(1+|z|^{2})}{\bf 1}_{2}
+
\frac{1}{1+|z|^{2}}
\left(
\begin{array}{cc}
0        & -\dot{\bar{z}} \\
\dot{z} & 0
\end{array}
\right)
\left(
\begin{array}{cc}
1  & \bar{z} \\
-z & 1
\end{array}
\right)
+
i\dot{\phi}\ 
U
\left(
\begin{array}{cc}
1 &      \\
   & -1
\end{array}
\right)
U^{\dagger}  \\
&=&
\frac{1}{1+|z|^{2}}
\left(
\begin{array}{cc}
-\frac{\dot{z}\bar{z}-z\dot{\bar{z}}}{2} & -\dot{\bar{z}} \\
\dot{z} & \frac{\dot{z}\bar{z}-z\dot{\bar{z}}}{2}
\end{array}
\right)
+
i\dot{\phi}\ 
\frac{1}{1+|z|^{2}}
\left(
\begin{array}{cc}
1-|z|^{2} & 2\bar{z}    \\
2z         & -1+|z|^{2}
\end{array}
\right)  \\
&=&
\frac{1}{1+|z|^{2}}
\left(
\begin{array}{cc}
-\frac{\dot{z}\bar{z}-z\dot{\bar{z}}}{2}+i\dot{\phi}(1-|z|^{2}) & 
-\dot{\bar{z}}+2i\dot{\phi}\bar{z} \\
\dot{z}+2i\dot{\phi}z & 
\frac{\dot{z}\bar{z}-z\dot{\bar{z}}}{2}-i\dot{\phi}(1-|z|^{2})
\end{array}
\right)
\end{eqnarray*}
and from (\ref{eq:2-evolution-equation}) we have
\[
\left\{
\begin{array}{l}
\frac{i}{1+|z|^{2}}(\dot{z}+2i\dot{\phi}z)=v, \\
\frac{i}{1+|z|^{2}}\left\{\frac{\dot{z}\bar{z}-z\dot{\bar{z}}}{2}-i\dot{\phi}(1-|z|^{2})\right\}
=-h
\end{array}
\right.
\]
or
\[
\left\{
\begin{array}{l}
\dot{z}+2i\dot{\phi}z=-iv(1+|z|^{2}),\  
\dot{\bar{z}}-2i\dot{\phi}\bar{z}=i\bar{v}(1+|z|^{2}), 
\\
\frac{\dot{z}\bar{z}-z\dot{\bar{z}}}{2}-i\dot{\phi}(1-|z|^{2})=ih(1+|z|^{2}).
\end{array}
\right.
\]
From this it is not difficult to show
\[
\left\{
\begin{array}{l}
\dot{z}+2i\dot{\phi}z=-iv(1+|z|^{2}), \\
\dot{\phi}=-\frac{v\bar{z}+\bar{v}z+2h}{2}
\end{array}
\right.
\]
and finally we obtain
\begin{eqnarray}
\label{eq:2-solution-1}
&&\dot{z}-i(\bar{v}z^{2}+2hz-v)=0, \\
\label{eq:2-solution-2}
&&\dot{\phi}=-\frac{v\bar{z}+\bar{v}z+2h}{2}.
\end{eqnarray}
The equation (\ref{eq:2-solution-1}) is a Riccati equation and 
(\ref{eq:2-solution-2}) is a phase equation. This is our canonical 
form of (\ref{eq:2-evolution-equation}). We believe that our derivation 
is smarter than that of \cite{RU-1}.

\vspace{3mm}
A comment is in order.\ \ Unfortunately, it is impossible to solve 
(\ref{eq:2-solution-1}) exactly, so we need some numerical computational 
method.

\section{Three Level System}
In this section we treat the three level system ($n=3$) in detail, 
which is the aim of the paper. However, the calculation is not easy 
compared to the two level one.

The Hamiltonian is
\begin{equation}
\label{eq:3-Hamiltonian}
H=
\left(
\begin{array}{ccc}
h_{1}(t) & \bar{v}_{1}(t) & \bar{v}_{2}(t) \\
v_{1}(t) & h_{2}(t) & \bar{v}_{3}(t)         \\
v_{2}(t) & v_{3}(t) & h_{3}(t) 
\end{array}
\right);\quad h_{1}(t)+h_{2}(t)+h_{3}(t)=0
\end{equation}
from (\ref{eq:Hamiltonian}). From the fact
\[
SU(3)\approx \frac{SU(3)}{U(1)^{2}}\times U(1)^{2},
\]
$U(t)$ can be parametrized as
\begin{equation}
\label{eq:3-decomposition}
U(t)=U_{1}(Z(t))U_{2}(t)
\end{equation}
where
\begin{equation}
U_{1}(Z(t))
=
\left(
\begin{array}{ccc}
1 & -\frac{\bar{x}(t)+\bar{y}(t)z(t)}{\Delta_{1}} & \frac{\bar{x}(t)\bar{z}(t)-\bar{y}(t)}{\Delta_{2}} \\
x(t) & 1-\frac{x(t)(\bar{x}(t)+\bar{y}(t)z(t))}{\Delta_{1}} & -\frac{\bar{z}(t)}{\Delta_{2}}     \\
y(t) & z(t)-\frac{y(t)(\bar{x}(t)+\bar{y}(t)z(t))}{\Delta_{1}} & \frac{1}{\Delta_{2}}
\end{array}
\right)
\left(
\begin{array}{ccc}
\frac{1}{\sqrt{\Delta_{1}}} &   &                \\
     & \sqrt{\frac{\Delta_{1}}{\Delta_{2}}} &  \\
     &        & \sqrt{\Delta_{2}}
\end{array}
\right)
\end{equation}
and
\begin{equation}
\Delta_{1}=1+|x(t)|^{2}+|y(t)|^{2},\quad  \Delta_{2}=1+|z(t)|^{2}+|x(t)z(t)-y(t)|^{2}
\end{equation}
and
\begin{equation}
U_{2}(t)=
\left(
\begin{array}{ccc}
e^{i\phi_{1}(t)} &    &                                   \\
               & e^{i\phi_{2}(t)} &                       \\
               &     & e^{-i(\phi_{1}(t)+\phi_{2}(t))}
\end{array}
\right).
\end{equation}
This form is convenient in the latter calculation, though it is  
a variant of that of \cite{FO-2}, \cite{Fu-2}, \cite{DJ}.

\noindent
Moreover, for the latter convenience we set
\begin{eqnarray}
V(t)&=&
\left(
\begin{array}{ccc}
1 & -\frac{\bar{x}(t)+\bar{y}(t)z(t)}{\Delta_{1}} & \frac{\bar{x}(t)\bar{z}(t)-\bar{y}(t)}{\Delta_{2}} \\
x(t) & 1-\frac{x(t)(\bar{x}(t)+\bar{y}(t)z(t))}{\Delta_{1}} & -\frac{\bar{z}(t)}{\Delta_{2}}     \\
y(t) & z(t)-\frac{y(t)(\bar{x}(t)+\bar{y}(t)z(t))}{\Delta_{1}} & \frac{1}{\Delta_{2}}
\end{array}
\right),  \\
D_{\Delta}(t)&=&
\left(
\begin{array}{ccc}
\frac{1}{\sqrt{\Delta_{1}}} &   &                \\
     & \sqrt{\frac{\Delta_{1}}{\Delta_{2}}} &  \\
     &        & \sqrt{\Delta_{2}}
\end{array}
\right).
\end{eqnarray}

Then $U(t)$ in (\ref{eq:3-decomposition}) can be written as
\[
U(t)=V(t)D_{\Delta}(t)U_{2}(t).
\]

In the following we omit $t$ from all variables ($x(t)\rightarrow x$, etc) 
for simplicity. 
From the equation (\ref{eq:evolution-equation}) we have
\[
i\left(
\dot{V}D_{\Delta}U_{2}+V\dot{D}_{\Delta}U_{2}+VD_{\Delta}\dot{U}_{2}
\right)
=HVD_{\Delta}U_{2}
\]
and
\begin{equation}
\label{eq:transform}
i\left(
\dot{V}+V\dot{D}_{\Delta}D_{\Delta}^{-1}+V\dot{U}_{2}U_{2}^{-1}
\right)
=HV
\end{equation}
because $D_{\Delta}$ and $U_{2}$ are diagonal. This form is 
better as shown in the following calculation.

\vspace{3mm}
A comment is in order.\ \ From (\ref{eq:transform}) we have 
a clear form
\[
H=
i\left(
\dot{V}V^{-1}+V\dot{D}_{\Delta}D_{\Delta}^{-1}V^{-1}+
V\dot{U}_{2}U_{2}^{-1}V^{-1}
\right).
\]
However, we don't recommend readers to calculate the right hand side 
because the calculation becomes very complicated.

\vspace{3mm}
The calculation is truly complicated. 
The (11), (21) and (31)--components of the matrix equation (\ref{eq:transform}) 
read
\begin{eqnarray*}
i\dot{x}-\frac{i}{2}\frac{\dot{\Delta}_{1}}{\Delta_{1}}x-\dot{\phi}_{1}x&=&
v_{1}+h_{2}x+\bar{v}_{3}y, \\
i\dot{y}-\frac{i}{2}\frac{\dot{\Delta}_{1}}{\Delta_{1}}y-\dot{\phi}_{1}y&=&
v_{2}+v_{3}x+h_{3}y, \\
-\frac{i}{2}\frac{\dot{\Delta}_{1}}{\Delta_{1}}-\dot{\phi}_{1}&=&
h_{1}+\bar{v}_{1}x+\bar{v}_{2}y.
\end{eqnarray*}
From these equations we have
\begin{eqnarray}
\label{eq:x-equation}
i\dot{x}
&=&(v_{1}+h_{2}x+\bar{v}_{3}y)-(h_{1}+\bar{v}_{1}x+\bar{v}_{2}y)x  \nonumber \\
&=&v_{1}+(h_{2}-h_{1})x-\bar{v}_{1}x^{2}+\bar{v}_{3}y-\bar{v}_{2}xy
\end{eqnarray}
and
\begin{eqnarray}
\label{eq:y-equation}
i\dot{y}
&=&(v_{2}+v_{3}x+h_{3}y)-(h_{1}+\bar{v}_{1}x+\bar{v}_{2}y)y  \nonumber \\
&=&v_{2}+(h_{3}-h_{1})y-\bar{v}_{2}y^{2}+v_{3}x-\bar{v}_{1}xy
\end{eqnarray}
and
\begin{equation}
\label{eq:pre-phi-1-equation}
\dot{\phi}_{1}
=-\frac{1}{2}\frac{i\dot{\Delta}_{1}}{\Delta_{1}}-(h_{1}+\bar{v}_{1}x+\bar{v}_{2}y).
\end{equation}

Next, we must calculate the term
\[
i\dot{\Delta}_{1}=i\frac{d}{dt}{(1+|x|^{2}+|y|^{2})}
=
i(\dot{x}\bar{x}+x\dot{\bar{x}}+\dot{y}\bar{y}+y\dot{\bar{y}}).
\]
Then from (\ref{eq:x-equation}) and (\ref{eq:y-equation}) 
it is not difficult to show
\[
i\dot{\Delta}_{1}=
\{(h_{1}+v_{1}\bar{x}+v_{2}\bar{y})-(h_{1}+\bar{v}_{1}x+\bar{v}_{2}y)\}\Delta_{1},
\]
so that we finally obtain
\begin{equation}
\label{eq:phi-1-equation}
\dot{\phi}_{1}=
-\frac{(h_{1}+\bar{v}_{1}x+\bar{v}_{2}y)+(h_{1}+v_{1}\bar{x}+v_{2}\bar{y})}{2}
\end{equation}
from (\ref{eq:pre-phi-1-equation}).

\vspace{3mm}
The (12), (22) and (32)--components of the matrix equation (\ref{eq:transform}) 
read
\begin{small}
\begin{eqnarray*}
&&-i\frac{d}{dt}\left(\frac{\bar{x}+\bar{y}z}{\Delta_{1}}\right)
-i\frac{\bar{x}+\bar{y}z}{\Delta_{1}}\times 
\frac{1}{2}\left(\frac{\dot{\Delta}_{1}}{\Delta_{1}}-\frac{\dot{\Delta}_{2}}{\Delta_{2}}\right)
-\frac{\bar{x}+\bar{y}z}{\Delta_{1}}(-\dot{\phi}_{2}) \\
&&\quad
=-h_{1}\frac{\bar{x}+\bar{y}z}{\Delta_{1}}
+\bar{v}_{1}\left\{1-\frac{x(\bar{x}+\bar{y}z)}{\Delta_{1}}\right\}
+\bar{v}_{2}\left\{z-\frac{y(\bar{x}+\bar{y}z)}{\Delta_{1}}\right\}, \\
&{}& \\
&&-i\dot{x}\frac{\bar{x}+\bar{y}z}{\Delta_{1}}
-ix\frac{d}{dt}\left(\frac{\bar{x}+\bar{y}z}{\Delta_{1}}\right)
+i\left\{1-\frac{x(\bar{x}+\bar{y}z)}{\Delta_{1}}\right\}\times
\frac{1}{2}\left(\frac{\dot{\Delta}_{1}}{\Delta_{1}}-\frac{\dot{\Delta}_{2}}{\Delta_{2}}\right)
+\left\{1-\frac{x(\bar{x}+\bar{y}z)}{\Delta_{1}}\right\}(-\dot{\phi}_{2}) \\
&&\quad
=-v_{1}\frac{\bar{x}+\bar{y}z}{\Delta_{1}}
+h_{2}\left\{1-\frac{x(\bar{x}+\bar{y}z)}{\Delta_{1}}\right\}
+\bar{v}_{3}\left\{z-\frac{y(\bar{x}+\bar{y}z)}{\Delta_{1}}\right\}, \\
&{}& \\
&&i\dot{z}
-i\dot{y}\frac{\bar{x}+\bar{y}z}{\Delta_{1}}
-iy\frac{d}{dt}\left(\frac{\bar{x}+\bar{y}z}{\Delta_{1}}\right)
+i\left\{z-\frac{y(\bar{x}+\bar{y}z)}{\Delta_{1}}\right\}\times 
\frac{1}{2}\left(\frac{\dot{\Delta}_{1}}{\Delta_{1}}-\frac{\dot{\Delta}_{2}}{\Delta_{2}}\right)
+\left\{z-\frac{y(\bar{x}+\bar{y}z)}{\Delta_{1}}\right\}(-\dot{\phi}_{2}) \\
&&\quad
=-v_{2}\frac{\bar{x}+\bar{y}z}{\Delta_{1}}
+v_{3}\left\{1-\frac{x(\bar{x}+\bar{y}z)}{\Delta_{1}}\right\}
+h_{3}\left\{z-\frac{y(\bar{x}+\bar{y}z)}{\Delta_{1}}\right\}.
\end{eqnarray*}
\end{small}
From these equations it is not difficult to show
\begin{eqnarray}
\label{eq:z-equation}
i\dot{z}
&=&v_{3}+h_{3}z-(h_{2}+\bar{v}_{3}z)z+(xz-y)(\bar{v}_{1}+\bar{v}_{2}z) 
\nonumber \\
&=&v_{3}+(h_{3}-h_{2})z-\bar{v}_{3}z^{2}+(xz-y)(\bar{v}_{1}+\bar{v}_{2}z)
\end{eqnarray}
and
\[
\frac{i}{2}\left(\frac{\dot{\Delta}_{1}}{\Delta_{1}}-\frac{\dot{\Delta}_{2}}{\Delta_{2}}\right)
-\dot{\phi}_{2}=h_{2}+\bar{v}_{3}z-x(\bar{v}_{1}+\bar{v}_{2}z)
\]
or
\begin{equation}
\label{eq:pre-phi-2-equation}
\dot{\phi}_{2}=
\frac{1}{2}\left(\frac{i\dot{\Delta}_{1}}{\Delta_{1}}-\frac{i\dot{\Delta}_{2}}{\Delta_{2}}\right)
-(h_{2}+\bar{v}_{3}z)+x(\bar{v}_{1}+\bar{v}_{2}z).
\end{equation}
Since
\[
\frac{i\dot{\Delta}_{1}}{\Delta_{1}}=
(h_{1}+v_{1}\bar{x}+v_{2}\bar{y})-(h_{1}+\bar{v}_{1}x+\bar{v}_{2}y)
\]
the remaining one is to calculate $\dot{\Delta}_{2}/\Delta_{2}$. However, 
it is very troublesome.

From
\begin{eqnarray*}
i\dot{\Delta}_{2}
&=&i\frac{d}{dt}{(1+|z|^{2}+|xz-y|^{2})} \\
&=&i\left\{
\dot{z}\bar{z}+z\dot{\bar{z}}+\frac{d}{dt}(xz-y)\overline{(xz-y)}+
(xz-y)\frac{d}{dt}\overline{(xz-y)}\right\}
\end{eqnarray*}
long calculation gives 
\[
\frac{i\dot{\Delta}_{2}}{\Delta_{2}}
=
-\bar{v}_{3}z+v_{3}\bar{z}+\bar{v}_{2}(xz-y)-v_{2}(\bar{x}\bar{z}-\bar{y})
\]
by use of (\ref{eq:x-equation}), (\ref{eq:y-equation}) and (\ref{eq:z-equation}). 
Therefore we finally obtain
\begin{equation}
\label{eq:phi-2-equation}
\dot{\phi}_{2}=
\frac{(-h_{2}-\bar{v}_{3}z+\bar{v}_{1}x+\bar{v}_{2}xz)+
(-h_{2}-v_{3}\bar{z}+v_{1}\bar{x}+v_{2}\bar{x}\bar{z})}{2}
\end{equation}
from (\ref{eq:pre-phi-2-equation}).

Let us summarize the result. Our canonical form is
\begin{eqnarray*}
i\dot{x}&=&v_{1}+(h_{2}-h_{1})x-\bar{v}_{1}x^{2}+\bar{v}_{3}y-\bar{v}_{2}xy, \\
i\dot{y}&=&v_{2}+(h_{3}-h_{1})y-\bar{v}_{2}y^{2}+v_{3}x-\bar{v}_{1}xy, \\
i\dot{z}&=&v_{3}+(h_{3}-h_{2})z-\bar{v}_{3}z^{2}+(xz-y)(\bar{v}_{1}+\bar{v}_{2}z)
\end{eqnarray*}
and
\begin{eqnarray*}
\dot{\phi}_{1}
&=&-\frac{(h_{1}+\bar{v}_{1}x+\bar{v}_{2}y)+(h_{1}+v_{1}\bar{x}+v_{2}\bar{y})}{2}, \\
\dot{\phi}_{2}
&=&
\frac{(-h_{2}-\bar{v}_{3}z+\bar{v}_{1}x+\bar{v}_{2}xz)+
(-h_{2}-v_{3}\bar{z}+v_{1}\bar{x}+v_{2}\bar{x}\bar{z})}{2}
\end{eqnarray*}
where $h_{1}+h_{2}+h_{3}=0$. 

\vspace{3mm}
A comment is in order.\ \ We have a set of complex Riccati differential 
equations interacting with one another (\cite{FO-2}) and real phase 
equations. Unfortunately, to solve them exactly is impossible, so we must 
use some numerical computational method.

\section{Discussion}
In this paper we obtained a canonical form of the evolution operator 
of a time--dependent Hamiltonian in the three level system ($n=3$), 
namely three complex Riccati differential equations interacting with 
one another and two real phase equations. 
For the case of $n\geq 4$ it is very hard to calculate, so we will 
leave calculation to (young) readers. 

The result looks good and may be applied to a wide class of problems 
in Quantum Physics\footnote{the author expects the result to solve 
important problems in Quantum Computation} or Mathematical Physics. 
Further work will be needed.


\end{document}